\documentclass[preprint,aps,amsmath,superscriptaddress,nofootinbib]{revtex4}
\usepackage{bm}
\usepackage{epsfig}

\newif\ifpdf
\ifx\pdfoutput\undefined
\pdffalse 
\else
\pdfoutput=1 
\pdftrue
\fi

\def\Dsl{\hbox{/\kern-.6000em D}} 

\def\dsl{\,\raise.15ex\hbox{/}\mkern-13.5mu D}

\def\ltap{\ \raise.3ex\hbox{$<$\kern-.75em\lower1ex\hbox{$\sim$}}\ }
\def\gtap{\ \raise.3ex\hbox{$>$\kern-.75em\lower1ex\hbox{$\sim$}}\ }
\def\OMIT#1{}

\def\vsl{v\!\!\!\slash}

\def\OMIT#1{}

\newcommand{\nn}{\nonumber}

\newcommand{\bea}{\begin{eqnarray}}
\newcommand{\eea}{\end{eqnarray}}

\begin{document}

\ifpdf
\DeclareGraphicsExtensions{.pdf, .jpg}
\newcommand{\picspace}{\vspace{-2.5in}}
\newcommand{\picspacehalf}{\vspace{-1.75in}}
\else
\DeclareGraphicsExtensions{.eps, .jpg,.ps}
\newcommand{\picspace}{\vspace{0in}}
\newcommand{\picspacehalf}{\vspace{0in}}
\fi



\title{Chiral Lagrangian with Heavy Quark-Diquark Symmetry} 

\author{Jie Hu\footnote{Electronic address: hujie@phy.duke.edu}}
\affiliation{Department of Physics, Duke University, Durham,  NC 27708\vspace{0.2cm}}

\author{Thomas Mehen\footnote{Electronic address: mehen@phy.duke.edu}}
\affiliation{Department of Physics, Duke University, Durham,  NC 27708\vspace{0.2cm}}
\affiliation{Jefferson Laboratory, 12000 Jefferson Ave., Newport News, VA 23606\vspace{0.2cm}}

\date{\today\\ \vspace{1cm} }



\begin{abstract}

We construct a chiral Lagrangian for doubly heavy baryons and heavy 
mesons that is invariant under heavy quark-diquark symmetry at leading order
and includes the leading $O(1/m_Q)$ symmetry violating operators. The theory is used to predict
 the electromagnetic decay width of the $J=\frac{3}{2}$ member 
of the ground state doubly heavy baryon doublet. Numerical estimates are provided for doubly charm baryons.
We also calculate chiral corrections to doubly heavy baryon masses and 
strong decay widths of low lying excited doubly heavy baryons.

\end{abstract}
\maketitle

\newpage

Heavy quark-diquark symmetry relates mesons with a single heavy quark to antibaryons with  two heavy antiquarks. Savage and Wise~\cite{Savage:di} argued
that quark-diquark symmetry was realized in the heavy quark limit of Quantum Chromodynamics (QCD) and studied this symmetry using the methods of Heavy Quark
Effective Theory (HQET)~\cite{Manohar:2000dt}. Recently, Refs.~\cite{Brambilla:2005yk,Fleming:2005pd} derived effective Lagrangians for heavy  diquarks
within the framework of Non-Relativistic QCD (NRQCD)~\cite{Bodwin:1994jh,Brambilla:1999xf,Luke:1999kz}. These papers obtain a prediction for the hyperfine
splitting of the ground state  doubly heavy baryons in terms of the ground state heavy  meson  hyperfine splitting\footnote{The formula for the hyperfine
splittings in Ref.~\cite{Savage:di} differs from the correct formula in Refs.~\cite{Brambilla:2005yk,Fleming:2005pd} by a factor of 2.}. Heavy
quark-diquark symmetry also relates other properties of heavy mesons and doubly heavy baryons. A useful tool for studying low energy strong and
electromagnetic interactions  of heavy hadrons  is heavy hadron chiral perturbation theory (HH$\chi$PT)~\cite{Wise:1992hn,Burdman:1992gh,Yan:1992gz}. This
theory has heavy hadrons, Goldstone bosons, and photons as its elementary degrees of freedom and incorporates the approximate chiral and heavy
quark symmetries of QCD.  In this paper we derive a  version of HH$\chi$PT that includes doubly heavy baryons and incorporates  heavy quark-diquark
symmetry.  The theory is used to calculate chiral corrections to doubly heavy baryon masses and to obtain model-independent predictions for the
electromagnetic decay of the $J=\frac{3}{2}$ member of the ground state doubly heavy baryon doublet. Our formulae are applicable to either doubly bottom or
doubly charm baryons, and we give numerical estimates for the case of doubly charm baryons. We also discuss the low lying excited 
doubly heavy baryons, show how these states can be included in the effective theory, and calculate their strong decay widths.

Motivation for this work comes from the SELEX experiment's recent observation  of states which have been  tentatively interpreted as doubly charm
baryons~\cite{Mattson:2002vu,Moinester:2002uw,Ocherashvili:2004hi}, and also the COMPASS experiment, which in its second phase run in 2006 hopes to
observe doubly charm baryons~\cite{Schmitt:2003gi}. Many aspects of the SELEX states are difficult to understand. States observed by SELEX include
the   $\Xi_{cc}^+(3520)$, which  decays weakly into $\Lambda_c^+ \pi^+ K^-$~\cite{Mattson:2002vu} as well as $p D^+
K^-$~\cite{Ocherashvili:2004hi}, the $\Xi_{cc}^{++}(3460)$, which decays weakly into $\Lambda_c^+ K^- \pi^+\pi^+$~\cite{Moinester:2002uw}, and a
broader state, $\Xi_{cc}^{++}(3780)$,  also seen to decay into $\Lambda_c^+ K^- \pi^+ \pi^+$~\cite{Moinester:2002uw}. The ground states of the
$\Xi^+_{cc}$ and $\Xi^{++}_{cc}$ are related by isospin symmetry and therefore should differ in mass by only a few MeV, so the observed difference
of 60 MeV seems implausible. On the other hand,  an unpublished talk~\cite{russ} and conference proceedings~\cite{Russ:2004vn} present evidence for
additional states, $\Xi_{cc}^{+}(3443)$ and $\Xi_{cc}^{++}(3541)$. If these states  exist the isospin splittings are closer to theoretical
expectations, but still quite large. The difference between the mass of the $\Xi_{cc}^+(3520)$ and $\Xi_{cc}^{+}(3443)$ is  77 MeV, and the
splitting between $\Xi_{cc}^{++}(3541)$ and $\Xi_{cc}^{++}(3460)$ is 81 MeV. These splittings agree remarkably well with  calculations of the
doubly charm hyperfine splittings in quenched lattice QCD~\cite{Lewis:2001iz,Mathur:2002ce,Flynn:2003vz} and are within $\sim 25$ MeV of the heavy
quark-diquark symmetry prediction obtained in Refs.~\cite{Brambilla:2005yk,Fleming:2005pd}, an acceptable discrepancy given the expected
$O(\Lambda_{\rm QCD}/m_c)$ corrections. However, interpretation of the $\Xi_{cc}^+(3520)$ as the $J=\frac{3}{2}$ member of the ground state doublet
is impossible to reconcile with the fact that  the
$\Xi^+_{cc}(3520)$ is observed to decay weakly because if the $\Xi_{cc}^{+}(3520)$ is not the ground state of the $ccd$ system it
should decay electromagnetically. There are also discrepancies between  the  weak decay lifetimes predicted by
HQET~\cite{Likhoded:1999yv,Guberina:1999mx,Kiselev:1998sy}($\sim$ 100 fs) and the  observed lifetimes ($<$ 33
fs)~\cite{Mattson:2002vu,Ocherashvili:2004hi}. Production cross sections are also poorly  understood within perturbative QCD~\cite{Kiselev:2002an}.
However, the SELEX states  are observed  in the forward region, $\langle x_F \rangle \sim 0.3$, where nonperturbative production mechanisms such as
intrinsic charm~\cite{Vogt:1995fs,Vogt:1995tf,Vogt:1994zf} or parton recombination~\cite{Braaten:2003vy,Braaten:2002yt} may be important.

Even if there is difficulty interpreting the SELEX data,  doubly  charm baryons must exist and are expected to have masses of approximately 3.5 GeV,
where the SELEX states are.  In light of existing and future experimental efforts to observe doubly charm baryons, it is desirable to have model
independent  predictions for other properties besides the relation for the hyperfine splittings derived in
Refs.~\cite{Savage:di,Brambilla:2005yk,Fleming:2005pd}. Therefore it is important to develop theoretical tools for analyzing the properties of doubly
heavy baryons systematically.  

Savage and Wise~\cite{Savage:di} wrote down a version of heavy quark effective theory (HQET) which
includes diquarks as elementary degrees of freedom and derived a formula relating heavy meson and doubly
heavy baryon hyperfine splittings. HQET only separates the scales $\Lambda_{\rm QCD}$ and $m_Q$, where
$m_Q$ is the heavy quark mass. The dynamics of a bound state of two heavy quarks is characterized by
additional scales $m_Q v$ and $m_Q v^2$, where $v$ is the typical velocity of the heavy quarks in the
bound state.  The correct effective theory for hadrons with two  heavy quarks is
Non-Relativistic QCD (NRQCD)~\cite{Bodwin:1994jh}, which properly accounts  for the scales $m_Q v$ and
$m_Q v^2$. Analysis of heavy diquarks within the framework of NRQCD was recently performed in
Refs.~\cite{Brambilla:2005yk,Fleming:2005pd}. These papers derived Lagrangians for diquark fields
starting from NRQCD and obtained the correct heavy quark symmetry prediction for the hyperfine splittings
of the doubly heavy baryons. For simplicity, we will consider only one flavor of heavy quark. 
The lowest mass diquark will consist of two heavy antiquarks in an orbital S-wave in the $\bf 3$ representation of color $SU(3)$.
Then Fermi statistics demands that they have total spin one. In the rest frame of the heavy quark and lowest mass
diquark, the  Lagrangian to $O(1/m_Q)$ is~\cite{Savage:di,Fleming:2005pd}
\bea \label{hqlag}
{\cal L} &=& h^\dagger \, \left( iD_0 - \frac{\vec{D}^2}{2 m_Q}\right) \, h +
\vec{V}^\dagger \, \cdot \left( iD_0 + \delta  - \frac{\vec{D}^2}{m_Q}\right) \, \vec{V}   
\nn \\ 
&& +\frac{g_s}{2 m_Q} h^\dagger \, \vec{\sigma}  \cdot {\vec B^a} \,
\frac{\lambda^a}{2}\, h 
+ \frac{i g_s}{2 m_Q} \vec{V}^\dagger \cdot  \vec{B}^a \,  \frac{\lambda^a}{2} \,\times \, \vec{V} 
\, .
\eea
Here $h$ is the heavy quark field, $\vec{V}$ is the field for the diquarks, the $\lambda^a/2$ are the $SU(3)$ color generators, 
${\rm Tr}[\lambda^a \lambda^b] = 2 \, \delta^{ab}$, $D_0$ and $\vec{D}$ are the time and spatial components of the gauge 
covariant derivative, respectively, $\vec{B}^a$ is the  chromomagnetic field, and $m_Q$ is the heavy quark mass. The term 
proportional to $\delta$ is the residual mass of the diquark. The heavy antiquarks in the diquark experience an
attractive force and therefore the mass of the  diquark is not $2 m_Q$ but $2 m_Q - \delta$, where
$\delta$ is the binding energy. This residual  mass can be removed by rephasing the diquark fields.
Physically, this corresponds to measuring diquark energies relative to the mass of the diquark, rather
than $2 m_Q$. Once this is done the Lagrangian, at lowest order in $1/m_Q$, is invariant under a $U(5)$
symmetry which  permutes the two spin states of the heavy quark and the three spin states of the heavy
antiquark. The $U(5)$ symmetry is broken by the $O(1/m_Q)$ kinetic energy and chromomagnetic couplings
of the heavy quark and diquark. The latter  terms are responsible for the hyperfine splittings.  

The ground state doublet of heavy mesons is usually represented in HH$\chi$PT  as a $4 \times 4$ matrix
transforming covariantly under Lorentz transformations, and transforming as a doublet under $SU(2)$ heavy quark 
spin symmetry,
\bea
H_v = \left(\frac{1+\vsl}{2}\right) ( P_v^{* \mu} \gamma_\mu - \gamma_5 P_v) \, .
\eea
Here $P^{* \mu}_v$ is the $J^P=1^-$ vector heavy meson field which  obeys the constraint $v_\mu P_v^{* \mu} = 0$,
where $v^\mu$ is the four-velocity of the heavy meson. $P_v$ is the $J^P=0^-$
pseudoscalar heavy meson field. The superfield $H_v$ obeys the constraints $\vsl H_v = - H_v \vsl = H_v$, so $H_v$ only 
has four independent degrees of freedom. These can be collected in a
$2 \times 2$ matrix. For example, in the heavy meson rest frame where $v^\mu =  (1,0,0,0)$,
\bea
H_{v} = \left( \begin{array}{cc} 0 & -\vec{P}^*_v \cdot \vec{\sigma} - P_v \\
                                  0 & 0 \end{array} \right) \, ,
\eea
where we have used the Bjorken-Drell conventions for $\gamma_\mu$ and $\gamma_5$. 
For a process such as the weak decay $B \to D \ell \nu$, in which the initial
and final heavy hadrons have different four-velocities, the covariant representation of fields
is needed to determine heavy quark symmetry constraints on heavy hadron form-factors.
However, for studying  low energy strong and electromagnetic interactions in which the heavy meson four-velocity is conserved 
(up to $O(\Lambda_{\rm QCD}/m_Q)$ corrections),
it is also possible to  work in the heavy meson rest frame and use $2 \times 2$ matrix fields. This makes some calculations
simpler and we find it easiest to formulate the extension of HH$\chi$PT with $U(5)$ quark-diquark
symmetry in this frame. We define the heavy meson field in our theory to be
\bea 
H_a = \vec{P}^*_a \cdot \vec{\sigma} + P_a  \, ,
\eea
where $a$ is an $SU(3)$ flavor anti-fundamental index and the $\vec{\sigma}$ are the Pauli matrices. Since we have chosen
to work in the heavy meson rest frame, Lorentz covariance is lost and the symmetries of the theory are rotational invariance, $SU(2)$ heavy quark spin
symmetry, parity, time reversal and $SU_L(3)\times SU_R(3)$ chiral symmetry. Under these symmetries the field $H_a$ transforms as
\bea
{\rm rotations} \qquad H_a^\prime &=& U H_a  U^\dagger \nn \\
{\rm heavy \,quark \,spin} \qquad H_a^\prime &=& S H_a  \nn \\
{\rm parity} \qquad  H_a^\prime &=&  - H_a \nn \\
{\rm time \, reversal} \qquad  H_a^\prime &=&  -\sigma_2 \, H_a^* \, \sigma_2\nn \\
SU_L(3)\times SU_R(3) \qquad  H_a^\prime &=&  H_b V_{ba}^\dagger \, .
\eea
Here $U$ and $S$ are $2 \times 2$ rotation matrices and  $V_{ba}^\dagger$ is an $SU(3)$ matrix which gives the 
standard nonlinear realization of $SU_L(3) \times SU_R(3)$ chiral symmetry. 
In the two component notation the HH$\chi$PT Lagrangian is:
\bea\label{hml}
{\cal L} = {\rm Tr}[H^\dagger_a (i D_0)_{ba} H_b] - g {\rm Tr}[H^\dagger_a  H_b \, \vec{\sigma}\cdot \vec A_{ba}]
+\frac{\Delta_H}{4}{\rm Tr}[H^\dagger_a \,  \sigma^i \, H_a \, \sigma^i] \, .
\eea
The last term breaks heavy quark spin symmetry and $\Delta_H$ is the hyperfine splitting of the heavy mesons.
 The time component of the covariant chiral derivative is $(D_0)_{ba}$, $\vec{A}_{ba}$ is the spatial part
of the axial vector field, and $g$ is the heavy meson axial coupling. Our definitions for the chiral covariant
derivative, the axial current, and the Lagrangian for the Goldstone boson fields are the same as Ref.~\cite{Stewart:1998ke}.

We are now ready to generalize the Lagrangian to incorporate the doubly heavy baryons and the $U(5)$ quark-diquark symmetry.
The field $H_a$ transforms like the tensor product of a heavy quark spinor and a light antiquark spinor.
(This is how representations of heavy hadron fields were constructed in Ref.~\cite{Falk:1991nq}.) 
Writing
the field with explicit indices, $(H_a)_{\alpha \beta}$, the index $\alpha$ corresponds to the spinor index of the heavy quark 
and the index $\beta$ is that of the light antiquark spinor. In the theory with quark-diquark symmetry,
the heavy quark spinor is replaced with a five-component field, the first two components corresponding to the 
two heavy quark spin states and the last three components corresponding to the three spin states of the diquark:
\bea 
{\cal Q}_\mu  = \left(\begin{array}{c} h_\alpha \\ V_i \end{array} \right) \, .
\eea 
In terms of ${\cal Q}_\mu$  the kinetic terms of the Lagrangian in Eq.(\ref{hqlag}) are
\bea
{\cal L} =  {\cal Q}_\mu^\dagger i D_0 {\cal Q}_\mu \, .
\eea
The fields in HH$\chi$PT with heavy quark-diquark symmetry transform as tensor products 
of the five component field ${\cal Q}_\mu$ and a two-component light antiquark spinor.
Thus, the $2 \times 2$ matrix  field $H_a$ is promoted to a $5 \times 2$ matrix field
\bea
H_{a,\alpha \beta} \rightarrow {\cal H}_{a,\mu \beta} = H_{a,\alpha \beta} + T_{a,i \beta} \, .
\eea
Here the index $\mu$ takes on values between 1 and 5, $\alpha,\beta = 1$ or 2,  and $i = 3,4$, or $5$. 
The doubly heavy baryon fields are contained in  $T_{a,i \beta}$. Under the symmetries of the theory 
${\cal H}_a$  transforms as
\bea
{\rm rotations} \qquad {\cal H}_a^\prime &=& {\cal R} {\cal H}_a  U^\dagger \nn \\
{\rm heavy \,quark \,spin} \qquad {\cal H}_a^\prime &=& S {\cal H}_a  \nn \\
{\rm parity} \qquad  {\cal H}_a^\prime &=&  - {\cal H}_a \nn \\
{\rm time \, reversal} \qquad  {\cal H}_a^\prime &=&  -\Sigma_2 \, {\cal H}_a^* \,\sigma_2 \nn \\
SU_L(3)\times SU_R(3)  \qquad  {\cal H}_a^\prime &=&  {\cal H}_b V_{ba}^\dagger 
\, .\eea 
The matrix $S$ is now an element of $U(5)$ and ${\cal R}$ is a $5 \times 5$ reducible rotation matrix
\bea
{\cal R}_{\mu \nu} = \left(\begin{array}{cc} U_{\alpha \beta} &  0  \\
                                        0  &   R_{ij} \end{array} \right) \,  ,
\eea
where $U_{\alpha \beta}$ is an $SU(2)$ rotation matrix and $R_{ij}$ is an orthogonal $3 \times 3$ rotation matrix  
related to $U$ by $U^\dagger \sigma_i U = R_{i j} \sigma_j$. The $5 \times 5$ matrix appearing in the time reversal
transformation is 
\bea
(\Sigma_2)_{\mu \nu} = \left(\begin{array}{cc} (\sigma_2)_{\alpha \beta} &  0  \\
                                        0  &   \delta_{ij} \end{array}\right)  \, .
\eea
Under rotations the field $T_{a,i\beta}$ transforms as
$T^\prime_{a,i \beta} = R_{ij}\, T_{a,j \gamma} \, U^\dagger_{\gamma \beta}$. $T_{a,i \beta}$ can be further 
decomposed into its spin-$\frac{3}{2}$ and spin-$\frac{1}{2}$ components,
\bea\label{tsf}
T_{a,i \beta} = \sqrt{2}\left(\Xi^*_{a,i \beta}+ \frac{1}{\sqrt{3}} \Xi_{a,\gamma} \,\sigma^i_{\gamma \beta} \right) \, ,
\eea
where $\Xi^*_{a,i \beta}$ and $\Xi_{a,\gamma}$ are the spin-$\frac{3}{2}$ and spin-$\frac{1}{2}$ fields, respectively.
The factor of $\sqrt{2}$ is a convention that ensures that the kinetic terms of the doubly heavy baryon fields have the same
normalization as the heavy meson fields. The field $\Xi^*_{a,i \beta}$ obeys the constraint  $\Xi^*_{a,i \beta} \, \sigma^i_{\beta
\gamma}=0$.  

The $U(5)$ invariant generalizations of the first two terms of Eq.~(\ref{hml}) are simply obtained by making the replacement
$H_a \rightarrow {\cal H}_a$. To determine the proper generalization of the $U(5)$ breaking term we note that the chromomagnetic
couplings in Eq.~(\ref{hqlag}) can be written as
\bea\label{chromo}
\frac{g_s}{2m_Q} {\cal Q}_\mu^\dagger \vec\Sigma_{\mu \nu} \cdot \vec{B^a} \frac{\lambda^a}{2}{\cal Q}_{\nu} \,  ,
\eea
where the $\vec{\Sigma}_{\mu \nu}$ are the $5 \times 5$ matrices
\bea
\vec{\Sigma}_{\mu \nu} = \left(\begin{array}{cc} \vec{\sigma}_{\alpha \beta} &  0  \\
                                        0  &   \vec{{\cal T}}_{jk} \end{array} \right) \,  ,
\eea
and $({\cal T}^i)_{jk} = -i \epsilon_{ijk}$. It is now obvious that the correct generalization of Eq.~(\ref{hml}) is
\bea\label{qdql}
{\cal L} &=& {\rm Tr}[{\cal H}^\dagger_a (i D_0)_{ba} {\cal H}_b] - g \, {\rm Tr}[{\cal H}^\dagger_a  {\cal H}_b \, \vec{\sigma}\cdot \vec A_{ba}]
+\frac{\Delta_H}{4}{\rm Tr}[{\cal H}^\dagger_a \,  \Sigma^i \, {\cal H}_a \, \sigma^i] \nn \\
&=&{\rm Tr}[H^\dagger_a (i D_0)_{ba} H_b] - g \,{\rm Tr}[H^\dagger_a  H_b \, \vec{\sigma}\cdot \vec A_{ba}]
+\frac{\Delta_H}{4}{\rm Tr}[H^\dagger_a \,  \sigma^i \, H_a \, \sigma^i] \nn \\
&&+{\rm Tr}[T^\dagger_a (i D_0)_{ba} T_b] - g \,{\rm Tr}[T^\dagger_a  T_b \, \vec{\sigma}\cdot \vec A_{ba}]
+\frac{\Delta_H}{4}{\rm Tr}[T^\dagger_a \,  {\cal T}^i \, T_a \, \sigma^i] \, .
\eea
The last line of Eq.~(\ref{qdql}) contains the terms relevant for doubly heavy baryons. Heavy quark-diquark symmetry relates the 
couplings in the doubly heavy baryon sector to the heavy meson sector. The propagator
for the spin-$\frac{1}{2}$ doubly heavy baryon is
\bea 
\frac{i \delta_{ab} \, \delta_{\alpha \beta} }{2(k_0 + \Delta_H/2 + i\epsilon)} \nn  \, , 
\eea
while the propagator for the spin-$\frac{3}{2}$ doubly heavy baryon is
\bea
\frac{i \delta_{ab} \, {\cal P}_{i\alpha,j \beta} }{2(k_0 - \Delta_H/4 + i\epsilon)}  =
\frac{i  \delta_{ab} \, (\delta_{ij}\delta_{\alpha \beta} -\frac{1}{3}(\sigma^i \sigma^j)_{\alpha \beta})}{2(k_0 - \Delta_H/4 + i\epsilon)}  
 \, . \nn
\eea
The projection operator ${\cal P}_{i\alpha,j \beta}$ satisfies $\sigma^i_{\gamma \alpha}\, {\cal P}_{i\alpha,j \beta} =
{\cal P}_{i\alpha,j \beta} \, \sigma^j_{\beta \gamma} = 0$. Comparison of the poles of the propagators shows that the hyperfine 
splitting for the doubly heavy baryons is $\frac{3}{4}\Delta_H$, reproducing the heavy quark-diquark symmetry prediction 
\bea\label{hfs}
m_{\Xi^*} - m_{\Xi} = \frac{3}{4}(m_{P^*} - m_P) \, ,
\eea
obtained in Refs.~\cite{Brambilla:2005yk,Fleming:2005pd}. 

We can also include operators which mediate electromagnetic decays. The Lagrangian for electromagnetic decays 
of the heavy mesons in the two-component notation is~\cite{Amundson:1992yp}
\bea
{\cal L} =   \frac{e \beta}{2}{\rm Tr}[H_a^\dagger H_b \,  \vec{\sigma} \cdot \vec{B} \, Q_{ab}] + 
\frac{e}{2 m_Q} Q^\prime{\rm Tr}[H_a^\dagger \vec{\sigma} \cdot \vec{B} H_a ] \, ,
\eea
where $Q_{ab} = {\rm diag}(2/3,-1/3,-1/3)$ is the light quark charge matrix, $\beta$ is the parameter introduced
in Ref.~\cite{Amundson:1992yp},
and $Q^\prime$  is the heavy quark charge. For charm, $Q^\prime = 2/3$. The first term is the magnetic moment coupling of the 
light degrees of freedom and the  second term is the 
magnetic moment coupling of the heavy quark. Both terms are needed  to understand
the observed electromagnetic branching fractions of the $D^{*+}$ and $D^{*0}$~\cite{Amundson:1992yp}.
The magnetic couplings of the heavy quark and diquark are 
\bea \label{emlag}
{\cal L}_{em} &=& \frac{e}{2 m_Q} Q^\prime h^\dagger \, \vec{\sigma}  \cdot {\vec B} \, h 
- \frac{i e}{m_Q} Q^\prime \, \vec{V}^\dagger \cdot  \vec{B}  \times \, \vec{V} \nn \\
&=& \frac{e}{2m_Q} \, Q^\prime \, {\cal Q}_\mu^\dagger \, \vec\Sigma_{\mu \nu}^\prime \cdot \vec{B} {\cal Q}_\nu \, ,
\eea
where the $\vec{\Sigma}_{\mu \nu}^{\prime}$ are the $5 \times 5$ matrices 
\bea
\vec{\Sigma}^{\prime}_{\mu \nu} = \left(\begin{array}{cc} \vec{\sigma}_{\alpha \beta} &  0  \\
                                        0  &  -2 \vec{{\cal T}}_{jk} \end{array} \right) \,  .
\eea
The  magnetic   coupling of the diquark has the opposite sign as that of the heavy quark 
because it is composed of two heavy antiquarks. The coefficient
of the chromomagnetic coupling of the diquark in Eqs.~(\ref{hqlag},\ref{chromo}) is a factor of 2 smaller than the coefficient
of the electromagnetic coupling of the diquark in Eq.~(\ref{emlag}) due to a color factor. The magnetic couplings in the  HH$\chi$PT Lagrangian for heavy
mesons and doubly heavy baryons are  
\bea
{\cal L} =   \frac{e \beta}{2}{\rm Tr}[{\cal H}_a^\dagger \,{\cal H}_b \,  \vec{\sigma} \cdot \vec{B}\, Q_{ab}] + 
\frac{e}{2 m_Q} Q^\prime{\rm Tr}[{\cal H}_a^\dagger \, \vec{\Sigma}^\prime \cdot \vec{B}\, {\cal H}_b ] \, .
\eea
The part of this Lagrangian involving the doubly heavy baryon fields is 
\bea
{\cal L} =   \frac{e \beta}{2}{\rm Tr}[T_a^\dagger \, T_b \,  \vec{\sigma} \cdot \vec{B}\, Q_{ab}] -
\frac{e}{m_Q} Q^\prime{\rm Tr}[T_a^\dagger \, \vec{\cal T} \cdot \vec{B} \, T_b ] \, .
\eea
This can be used to obtain the following tree level predictions for the electromagnetic decay widths:
\bea \label{emdecay}
\Gamma[P_a^* \rightarrow P_a \gamma] &=& \frac{\alpha}{3} \left( \beta Q_{aa} +\frac{Q^\prime}{m_Q} \right)^2 
\,\frac{m_P}{m_{P^*}} E_\gamma^3 \nn \\
\Gamma[\Xi_a^*\rightarrow \Xi_a \gamma] &=&\frac{4 \alpha}{9} \left( \beta Q_{aa} -\frac{Q^\prime}{m_Q} \right)^2 \, 
\frac{m_{\Xi}}{m_{\Xi^*}} E_\gamma^3 \, .
\eea
Here $E_\gamma$ is the photon energy. These results could also be obtained in the quark model, with the parameter
$\beta=1/m_q$, where $m_q$ is the light constituent quark mass. The effective theory allows one to improve upon this 
approximation by including corrections from loops with light Goldstone bosons, which give $O(\sqrt{m_q})$ corrections
to the decay rates~\cite{Amundson:1992yp}. If these loop corrections are evaluated in an approximation where heavy hadron
mass differences are neglected, the correction to the above formulae can be incorporated by making the following 
replacements \cite{Amundson:1992yp}
\bea \label{su3}
\beta Q_{11} &\rightarrow& \frac{2}{3}\beta -\frac{g^2 m_K}{4 \pi f_K^2} -\frac{g^2 m_\pi}{4 \pi f_\pi^2}  \nn \\
\beta Q_{22} &\rightarrow&   -\frac{1}{3}\beta +\frac{g^2 m_\pi}{4 \pi f_\pi^2}   \nn \\
\beta Q_{33} &\rightarrow&  -\frac{1}{3}\beta +\frac{g^2 m_K}{4 \pi f_K^2}  \, .
\eea
For charm mesons, hyperfine splittings are $\approx 140$ MeV and the $SU(3)$ splitting is $\approx 100$ MeV,
while for bottom mesons the hyperfine splittings are $\approx 45$ MeV and  $SU(3)$ splitting is $\approx 90$ MeV.
The approximation of neglecting heavy hadron  mass differences and keeping Goldstone boson masses is reasonable for 
kaon loops but not for loops with pions. However, the largest  $O(\sqrt{m_q})$ corrections come from loops with kaons.
When data on double heavy baryon electromagnetic decays is available,
more accurate calculations along the lines of Ref.~\cite{Stewart:1998ke} should be performed. In this paper,
we will use Eqs.~(\ref{emdecay}) and (\ref{su3}) to obtain estimates of doubly charm baryon electromagnetic decay widths. 

Currently $\Gamma[D^{*+}]$ is measured to be $96 \pm 22$ keV, while there is only an upper limit 
for $\Gamma[D^{*0}]$. The branching ratios for the $D^{*+}$  decays are ${\rm Br}[D^{*+}\to D^0 \pi^+] = 67.7 \pm 0.5$\%,  
${\rm Br}[D^{*+}\to D^+ \pi^0] = 30.7 \pm 0.5$\% and ${\rm Br}[D^{*+}\to D^+ \gamma] = 1.6 \pm 0.4$\%. The branching ratios for $D^{*0}$  decays are
${\rm Br}[D^{*0}\to D^0 \pi^0] = 61.9 \pm 2.9$\% and ${\rm Br}[D^{*0}\to D^0 \gamma] = 38.1 \pm 2.9$\%.
Isospin symmetry can be used to relate the strong partial width of the $D^{*0}$   to 
 the known strong partial width of the $D^{*+}$. Then the measured branching fractions of the $D^{*0}$
can be used to obtain the partial electromagnetic width of the $D^{*0}$. We find 
\bea\label{emmes}
\Gamma[ D^{*0}\to D^0 \gamma] &=& 26.1 \pm 6.0 \, {\rm keV} \nn \\
\Gamma[ D^{*+}\to D^+ \gamma] &=& 1.54 \pm 0.35 \, {\rm keV} \, ,
\eea
where the error is dominated by the uncertainty in $\Gamma[D^{*+}]$. $\Gamma[ D^{*+}\to D^+ \gamma]$ is suppressed because
of a partial cancellation between the magnetic moments of the light degrees of freedom and the charm quark.
Using the partial widths in Eq.~(\ref{emmes}) and the formulae in Eqs.~(\ref{emdecay}) and (\ref{su3}), 
we obtain predictions for doubly charm baryon electromagnetic decays in Table \ref{table}.
\begin{table}[t]
\begin{tabular}{|c|c|c|c|c|}
\hline & & & & \\[-4 mm]\quad {\large Fit} \quad &\quad {\large $\beta^{-1}({\rm MeV})$} \quad  &
\quad {\large $m_c ({\rm MeV}) $}\quad & \qquad {\large $\Gamma[\Xi_{cc}^{*++} ] \,({\rm keV})$ } 
\qquad & \qquad {\large $\Gamma[\Xi_{cc}^{*+} ]\, (\rm{keV})$ }\qquad\\[-4mm]
 & &  & &\\
\hline   & &  & &\\[-4mm]
\quad QM 1 \quad  & 379  & 1863
 &   3.3 {\large$\left(\frac{E_\gamma}{80\, {\rm MeV}}\right)^3$ }&  2.6 {\large  $\left(\frac{E_\gamma}{80 \,{\rm MeV}}\right)^3$ }\\[5mm]
\quad QM 2 \quad&  356  & 1500
 &    3.4 {\large$\left(\frac{E_\gamma}{80 \,{\rm MeV}}\right)^3$ }&  3.2 {\large $\left(\frac{E_\gamma}{80 \,{\rm MeV}}\right)^3$ }\\[5mm]
\quad $\chi$PT 1 \quad & 272   &  1432 
 &  2.3    {\large$\left(\frac{E_\gamma}{80\, {\rm MeV}}\right)^3$ }   &   3.5 {\large $\left(\frac{E_\gamma}{80 \,{\rm MeV}}\right)^3$ }\\[5mm]
\quad $\chi$PT 2 \quad & 276  & 1500
 &   2.3    {\large$\left(\frac{E_\gamma}{80 \,{\rm MeV}}\right)^3$ }  &  3.3 {\large $\left(\frac{E_\gamma}{80 \, {\rm MeV}}\right)^3$  }\\[5mm]
\hline
\end{tabular}
\caption{Predictions for the electromagnetic widths of the $\Xi_{cc}^{*+}$ and $\Xi_{cc}^{*++}$. The fits are explained in the text.} \label{table}
\end{table}

In our calculations of the doubly charm baryon decay widths the factor $m_\Xi/m_{\Xi^*}$ in Eq.~(\ref{emdecay}) has been set equal to one. For the expected masses  and
hyperfine splittings of the doubly charm baryons, this factor changes the predictions for the widths by less than 3\%.  The fits are labeled in the left hand column of
Table \ref{table}. In the fits labeled QM we have not included the $O(\sqrt{m_q})$ corrections in Eq.~(\ref{su3}).  Therefore, these predictions for the doubly charm
baryon electromagnetic decays are the same as what would be obtained in the quark model. The values  of the parameters $\beta$ and $m_c$ for each fit are shown along
with the predictions for the electromagnetic decay widths. In QM 1, we have treated $\beta$ and $m_c$ as free parameters and fit these to the central values in
Eq.~(\ref{emmes}). In QM 2 we have set $m_c = 1500$ MeV and performed a least squared fit to $\beta$. In the fits labeled $\chi$PT, we have included the leading
$O(\sqrt{m_q})$ chiral corrections in Eq.~(\ref{su3}). We have used  $f_\pi=130$ MeV, $f_K = 159$ MeV, and  $g = 0.6$ which is extracted from a tree level fit to the
$D^{*+}$ width. In $\chi$PT  1, we fixed  $\beta$ and $m_c$ to reproduce the central values in Eq.~(\ref{emmes}). In $\chi$PT  2, we set $m_c = 1500$ MeV and performed
a least squares fit to $\beta$. There are several sources of error in the  calculation. We expect 30\% theoretical errors due to heavy quark symmetry breaking effects,
30\% errors due to higher order $SU(3)$ breaking effects, and 25\% uncertainty from the experimentally measured value of $\Gamma[D^{*+}]$ leading to at least 50\%
error in the predictions in Table~\ref{table}.

Chiral perturbation theory and the nonrelativistic quark model give similar size estimates for the $\Xi_{cc}^*$ electromagnetic decay widths which are
expected to be $\sim$2-3 keV if the hyperfine splitting is 80 MeV. The electromagnetic decay should completely dominate any possible weak
decay, even if the weak decay rates are an order of magnitude greater than calculated in Refs.~\cite{Likhoded:1999yv,Guberina:1999mx,Kiselev:1998sy}.  The quark model predicts
$\Gamma[\Xi_{cc}^{*++}]$ slightly greater than $\Gamma[\Xi_{cc}^{*+}]$. This is in contrast with the charm meson sector where the
magnetic moment of the light degrees of freedom
and the magnetic moment of the charm quark add constructively  to give a large  $\Gamma[D^{*0}\to D^0\gamma]$ and destructively to give a small $\Gamma[D^{*+}\to D^+
\gamma]$. In the doubly heavy baryon sector, the relative sign of the magnetic moments is reversed, and both decay rates are approximately the same. In fact
from Eq.~(\ref{emdecay}), we can  see that for $\beta = 4/m_c$ the two rates are exactly equal in the quark model. Fits to the $D^*$ electromagnetic decays
yield values of $\beta$ and $m_c$ that are close to this point in parameter space. Including the $O(\sqrt{m_q})$ corrections from chiral perturbation
theory, the most important effect is the kaon loop correction whose contribution to the $\Xi_{cc}^{*++}$ decay has opposite sign as the contribution from
$\beta$ at tree level, therefore suppressing the $\Xi_{cc}^{*++}$ decay relative to $\Xi_{cc}^{*+}$.

The theory can also be used to compute chiral corrections to doubly heavy baryon masses. The one loop corrections
to the hadron masses  are
\bea \label{mass}
\delta m_{\Xi_a^*} &=& \sum_{i,b} {\cal C}^i_{ab} \frac{g^2}{16 \pi^2 f_i^2} \left( \frac{5}{9} K(m_{\Xi^*_b} -m_{\Xi^*_a},m_i,\mu) +
\frac{4}{9} K(m_{\Xi_b} -m_{\Xi^*_a},m_i,\mu) \right)  \nn \\
 \delta m_{\Xi_a} &=& \sum_{i,b} {\cal C}^i_{ab} \frac{g^2}{16 \pi^2 f_i^2} \left( \frac{1}{9} K(m_{\Xi_b} -m_{\Xi_a},m_i,\mu) +
\frac{8}{9} K(m_{\Xi^*_b} -m_{\Xi_a},m_i,\mu) \right) \nn \\
\delta m_{H_a} &=& \sum_{i,b} {\cal C}^i_{ab} \frac{g^2}{16 \pi^2 f_i^2}   K(m_{H^*_b} -m_{H_a},m_i,\mu)    \nn \\
 \delta m_{H^*_a} &=& \sum_{i,b} {\cal C}^i_{ab} \frac{g^2}{16 \pi^2 f_i^2}  \left( \frac{1}{3} K(m_{H_b} -m_{H^*_a},m_i,\mu) +
\frac{2}{3} K(m_{H^*_b} -m_{H^*_a},m_i,\mu) \right)  \, .
\eea
Here $m_i$ and $f_i$ are the mass and decay constant of the Goldstone boson in the one loop diagram and ${\cal C}^i_{ab}$ is a factor which comes
from $SU(3)$ Clebsch-Gordan coefficients in the couplings. 
For loops with charged pions we have
 ${\cal C}^{\pi^{\pm}}_{12} = {\cal C}^{\pi^{\pm}}_{21} =1$, for loops with neutral pions
${\cal C}^{\pi^0}_{11} = {\cal C}^{\pi^0}_{22} = \frac{1}{2}$, for loops with kaons
  ${\cal C}^{K}_{3i} = {\cal C}^{K}_{i3} =1$ (i =1 or 2), and for loops with $\eta$ mesons
 $C^\eta_{11}=C^\eta_{22}= \frac{1}{6}$ and $C^\eta_{33} =\frac{2}{3}$.  The function $K(\delta,m,\mu)$ is 
 related to the finite part of the integral
\bea 
i \int \frac{d^D l}{(2 \pi)^D} \frac{\vec{l}^2}{l^2 - m_\pi^2 +i \epsilon}\frac{1}{l_0 - \delta + i \epsilon}=
\frac{1}{(4\pi)^2} K(\delta,m,\mu) \, ,
\eea
evaluated using dimensional regularization in the $\overline{MS}$ scheme. We find
\bea 
K(\delta,m,\mu) =    (- 2 \, \delta^3 + 3 \,m^2 \, \delta)\, \ln\left(\frac{m^2}{\mu^2}\right)
+ 2\, \delta \, (\delta^2-m^2)\, F\left(\frac{\delta}{m}\right) + 4 \,\delta^3-5 \,\delta \, m^2   \, ,
\eea
where
\begin{eqnarray}
F(x)  &=&  2\frac{\sqrt{1-x^2}}{x}\left[\frac{\pi}{2} - {\rm Tan}^{-1}\left(\frac{x}{\sqrt{1-x^2}}\right) \right]
\qquad  \quad |x| < 1 \nn \\
&=&  -2\frac{\sqrt{x^2-1}}{x} \, \ln \left(x+\sqrt{x^2 - 1} \,\right)  
\qquad  \qquad \quad \,\, \,|x| > 1 \, , \nn
\end{eqnarray}
and $\mu$ is the renormalization scale. The $\mu$ dependence in the one loop calculation is cancelled by counterterms that 
have not been included.

We are interested in how the one loop corrections affect the leading order prediction for the 
hyperfine splittings. Unfortunately, it is impossible to give a reliable estimate without knowing the 
numerical value of the counterterms required to cancel the $\mu$ dependence in the nonanalytic contribution.
Furthermore, to compute the contribution from kaon loops, one must know the masses of doubly charm strange baryons
which have not been observed. We will assume that the ground state doubly charm strange baryons are 100 MeV higher in mass
than their nonstrange counterparts, similar to the $D$ meson system. 
We work in the isospin limit and use $g=0.6$, $\Delta_H=140$ MeV, $m_\pi = 137$ MeV, $m_K=496$ MeV, $m_\eta = 548$ MeV and the
experimental values of the pseudoscalar meson decay constants:
 $f_\pi = 130$ MeV, $f_K = 159$ MeV, and $f_\eta$ = 156 MeV.
The nonanalytic part of the one loop correction to the nonstrange doubly charm baryon hyperfine splitting is
\bea
 \delta m_{\Xi_{cc}^*} - \delta m_{\Xi_{cc}}\, = \left\{
\begin{array}{cc}
\quad -7.0 \, {\rm MeV} & \quad \mu = 500 \, {\rm MeV} \\
\quad 8.1 \,{\rm MeV} & \quad \mu = 1000 \, {\rm MeV}\\
\quad 16.9 \, {\rm MeV} & \quad \mu = 1500 \, {\rm MeV} 
\end{array} \right. \, ,
\eea 
where we have shown our results for three values of $\mu$. For these choices of $\mu$ the nonanalytic part of the  chiral correction varies between
-7 MeV and +17 MeV. The nonanalytic part of the chiral correction to the doubly charm baryon hyperfine splitting is quite sensitive
to the choice of $\mu$, and lies within
$15$\% of the tree level prediction. We also calculate the correction to the hyperfine splitting  relationship of Eq.~(\ref{hfs}) and find for the
masses in the nonstrange sector
\bea\label{dhfs}
 \delta m_{\Xi_{cc}^*} - \delta m_{\Xi_{cc}} - \frac{3}{4}(\delta   m_{D^*} - \delta   m_D) = \left\{
\begin{array}{cc}
\quad 3.9 \, {\rm MeV} & \quad \mu = 500 \, {\rm MeV} \\
\quad 5.3 \,{\rm MeV} & \quad \mu = 1000 \, {\rm MeV}\\
\quad 6.1 \, {\rm MeV} & \quad \mu = 1500 \, {\rm MeV} 
\end{array} \right . \, .
\eea 
The nonanalytic correction to the symmetry prediction is small ($<$ 10 MeV) and relatively insensitive to the choice of $\mu$.
Chiral perturbation theory predicts the nonanalytic dependence
of the doubly heavy baryon masses on the light quark masses, and  generalized to 
include the effects of quenching as well as other lattice artifacts, formulae such as those
in Eq.~(\ref{mass}) should be useful for chiral extrapolations of doubly heavy baryon masses
and hyperfine splittings in lattice simulations.

Finally, we discuss excited doubly heavy baryons. There are two types of excitations in the doubly heavy baryon system: excitations of the light
degrees of freedom and excitations of the diquark. Excitations of the first type are related to analogous excitations in the heavy meson sector by
heavy quark-diquark symmetry. The lowest lying excited charm mesons  are in a doublet of $J^P=0^+$ and $1^+$ mesons with masses approximately 425
MeV above the ground state in the nonstrange sector~\cite{Anderson:1999wn,Abe:2003zm,Link:2003bd} and 350 MeV above the ground state in the strange
sector~\cite{Aubert:2003fg,Besson:2003jp}. In the nonstrange sector these states decay via S-wave pion emission and have widths in the range 
250-350 MeV, while in the strange sector the strong decay is via $\pi^0$ emission which violates isospin, and therefore the states are very narrow
with widths expected to be of order 10 keV~\cite{Mehen:2005hc}. These states have light degrees of freedom with  angular momentum and parity $j^p =
\frac{1}{2}^+$. The doubly charm baryons related to the even-parity excited charm mesons by quark-diquark symmetry  are a doublet  with $J^P=
\frac{1}{2}^+$ and $J^P= \frac{3}{2}^+$.  The excited charm mesons and doubly charm baryons can be incorporated into HH$\chi$PT with  a $5 \times
2$ matrix field ${\cal S}_{\mu \beta}$ which is like the field ${\cal H}_{\mu \beta}$ except  ${\cal S}_{\mu \beta}$ has opposite parity. The
excitation energies and strong decay widths of these excited doubly charm baryons should be similar to their counterparts in the charm meson
sector. Since the excited $\Xi^{++}_{cc}(3780)$ state observed by SELEX is only  320 MeV above the $\Xi^{++}_{cc}(3460)$, the lowest mass
$\Xi_{cc}^{++}$ candidate, and its width is considerably less than  300 MeV, it does not seem likely that this excited doubly charm baryon is
related to the excited charm mesons by heavy quark-diquark symmetry.

This is not unexpected as the lowest lying excited doubly charm baryons are not excitations of the light degrees of freedom but
rather states in which the diquark is excited. The lowest mass excited diquark is a  P-wave excitation. Because of Fermi statistics
the diquark is a heavy quark spin singlet. The diquark's orbital angular momentum couples with the angular momentum of the light
degrees of freedom to form  baryons with $J^P=\frac{1}{2}^+$ and $J^P=\frac{3}{2}^+$, which we will refer to as $\Xi_{cc}^{\cal P}$
and  $\Xi_{cc}^{{\cal P}*}$, respectively. The next lowest lying states are doubly heavy baryons with a radially excited diquark,
which form a heavy quark doublet with  $J^P=\frac{1}{2}^-$ and $J^P=\frac{3}{2}^-$ baryons, which we will refer to as
$\Xi_{cc}^{\prime}$ and  $\Xi_{cc}^{\prime *}$, respectively. If the heavy antiquarks are sufficiently heavy that the force between
them is approximately Coulombic, they interact via a potential which is 1/2 as strong as the potential between the quark and
antiquark in a quarkonium bound state.  Therefore we expect the excitation energies of the charm diquarks to be significantly  
smaller than the analogous excitation energies in charmonium. Quark model calculations of excited doubly charm baryons predict that 
the  $\Xi_{cc}^{{\cal P}}$ and $\Xi_{cc}^{{\cal P *}}$ states are about 225 MeV above the $\Xi_{cc}$ and $\Xi_{cc}^*$, respectively, and
that that the heavy quark doublet containing  $\Xi_{cc}^{\prime}$ and $\Xi_{cc}^{{\prime}*}$ is about  $300$ MeV above the ground state   
doublet~\cite{Ebert:2002ig,Kiselev:2002iy,Gershtein:2000nx,Gershtein:1998sx,Gershtein:1998un}.
These excitation energies are about 1/2 the corresponding excitation energies in the charmonium  system: $m_{h_c}-m_{J/\psi} = 430$
MeV and $m_{\psi^\prime}-m_{J/\psi} = 590$ MeV. The charm diquark excitation energies are less than the expected excitation energy of
the light degrees of freedom and therefore the lowest lying excited doubly charm baryons have excited diquarks. Excitation energies of a 
diquark made from two bottom quarks are similar to the excitation energies of a diquark made from charm, so the same conclusion holds 
for doubly bottom baryons.

The doubly heavy baryons with P-wave excited diquarks decay to the ground state via S-wave pion emission. These decays violate 
heavy quark spin symmetry because the total spin of the diquark is changed in the transition.
The Lagrangian for the excited $\Xi^{\cal P}$ and $\Xi^{\cal P *}$ states, including kinetic terms, residual mass
terms and terms which mediate the S-wave decays, is
\bea
{\cal L} &=& 2 \, (\Xi^{{\cal P}}_a)^\dagger \, \left( i \, (D_0)_{ba} - \delta_{{\cal P}}\, \delta_{ab}\right) \, \Xi^{\cal P}_{b}
+  2 \, (\Xi^{{\cal P} *}_{a})^\dagger \, \left( i \, (D_0)_{ba} - \delta_{{\cal P}*}\, \delta_{ab}\right) \,\Xi^{{\cal P}*}_{b}
\nn \\
&&+  2\, \lambda_{1/2} \, \left( \Xi^\dagger_{a } \,\Xi^{\cal P}_{b} A^0_{ba} +h.c. \, \right)
+ 2\, \lambda_{3/2}\,\left( \Xi^{*\dagger}_{a} \,\Xi^{\cal P *}_{b} A^0_{ba} +h.c.  \, \right) \, .
\eea
The strong decay widths of the P-wave excited nonstrange doubly charm baryons are 
\bea
\Gamma[\Xi_{cc}^{\cal P *} \to \Xi_{cc}^* \, \pi] &=& \frac{\lambda_{3/2}^2}{2 \pi f^2} \left(\frac{1}{2} E_{\pi^0}^2 p_{\pi^0}
+E_{\pi^+}^2 p_{\pi^+} \right)  \frac{m_{\Xi^*}}{m_{\Xi^{\cal P *}}} 
= \lambda_{3/2}^2 \, 111 \,{\rm MeV} \nn \\
\Gamma[\Xi_{cc}^{\cal P} \to \Xi_{cc} \, \pi] &=& \frac{ \lambda_{1/2}^2 }{2 \pi f^2} \left(\frac{1}{2} E_{\pi^0}^2 p_{\pi^0}
+E_{\pi^+}^2 p_{\pi^+} \right) \frac{m_{\Xi}}{m_{\Xi^{\cal P}}}
= \lambda_{1/2}^2 \,  111 \,{\rm MeV} \, .\nn  
\eea
To obtain numerical estimates, we have assumed the masses $m_{\Xi_{cc}} = 3440$ MeV, $m_{\Xi_{cc}^*} = 3520$ MeV, $m_{\Xi_{cc}^{\cal P}} = 3665$ MeV
and $m_{\Xi_{cc}^{\cal P*}} = 3745$ MeV, corresponding to a diquark excitation energy of 225 MeV.  We sum over both charged and neutral pion decay
modes.  The coupling constants $\lambda_{1/2}$ and $\lambda_{3/2}$ are $O(\Lambda_{\rm QCD}/m_Q)$ so we should  expect this suppression makes
$\lambda_{1/2}$ and $\lambda_{3/2} < 1$. Therefore these states could be quite narrow despite decaying via S-wave pion emission. The small widths are
due to the small excitation energy which leaves little phase space for the decay. If the excitation energy is increased to 280 MeV, the widths are  
twice as large. Like the isospin violating decays of the  $D_s^*$~\cite{Cho:1994zu} and the even-parity excited $D_s$ mesons~\cite{Colangelo:2003vg,Mehen:2005hc},
the excited doubly heavy strange baryons below the kaon threshold decay through a virtual $\eta$ which mixes into a $\pi^0$. Denoting the ground
state doubly charm strange baryons as $\Omega^{(*)}_{cc}$ and the P-wave excited doubly charm strange baryons as $\Omega_{cc}^{{\cal P} (*)}$
we obtain the following formulae for the isospin violating strong decay  widths 
\bea
\Gamma[\Omega_{cc}^{\cal P *} \to \Omega_{cc}^* \, \pi^0] &=& \frac{\lambda_{3/2}^2}{2 \pi f^2} \,\frac{2}{3} \, \theta^2\, E_{\pi^0}^2 p_{\pi^0}  \nn \\
\Gamma[\Omega_{cc}^{\cal P} \to \Omega_{cc} \, \pi^0] &=& \frac{ \lambda_{1/2}^2 }{2 \pi f^2}\,\frac{2}{3}\, \theta^2\, E_{\pi^0}^2 p_{\pi^0}
 \, .\nn  
\eea
Here $\theta=0.01$ is the $\pi^0-\eta$ mixing angle.
We expect these widths to be in the range 1-5 keV, but without knowing the masses of the $\Omega^{(*)}_{cc}$ and $\Omega_{cc}^{{\cal P} (*)}$
states or the couplings $\lambda_{1/2}$ and $\lambda_{3/2}$ we cannot make more precise predictions.

The $J^P=\frac{3}{2}^-$ and $J^P=\frac{1}{2}^-$ doubly heavy baryons with radially excited diquarks are members of a heavy quark doublet we will
denote $T^\prime_a$ whose definition in terms of component fields is identical to Eq.~(\ref{tsf}).  The Lagrangian describing  this field, including
terms which mediate its decay to the ground state, is
\bea\label{rexc}
{\cal L} &=& 
{\rm Tr}[T^{\prime \dagger}_a ( (i D_0)_{ab} - \delta_{T^\prime} \, \delta_{ab})  T^{\prime}_b] 
- g \, {\rm Tr}[T^{\prime \dagger}_a \,  T^{\prime}_b \, \vec{\sigma}\cdot \vec A_{ba}]
+\frac{\Delta_H}{4}{\rm Tr}[ T^{\prime \dagger}_a \,  {\cal T}^i \,  T^{\prime}_b \, \sigma^i] \nn \\
&& - \tilde{g}\,\left( {\rm Tr}[ T^{\dagger}_a \, T^{\, \prime}_b \, \vec{\sigma}\cdot \vec A_{ba}] + {\rm h.c.} \right) \, .
\eea
In the limit of infinite heavy quark mass, the light degrees of freedom in the radially excited doubly heavy baryons are in the same configuration as
the ground state. Therefore, they are also related to the heavy meson ground state doublet by heavy quark-diquark symmetry. The axial coupling  and
hyperfine splitting of $T^\prime_a$ are the same as $T_a$, as  long as the spatial extent of the excited diquark, which is of order $1/(m_Q v)$, is
much smaller than $1/\Lambda_{\rm QCD}$. This is valid in the heavy quark limit, but could receive significant corrections in the charm sector.  The
last term in Eq.~(\ref{rexc}) mediates P-wave decays from the excited $J^P=\frac{3}{2}^-$ and $J^P=\frac{1}{2}^-$  doubly heavy baryons 
to the ground state. The partial decay widths are
\bea\label{strong}
\Gamma[\Xi_a^{\prime *} \to \Xi_b^* \, \pi ]  &=&{\cal C}_{ab}\, \frac{5}{9} \, \frac{{\tilde g}^2}{2 \pi f^2} \,  \frac{m_{\Xi^*}}{m_{\Xi^{\prime *}}}|p_\pi|^3 \qquad
\Gamma[\Xi_a^{\prime *} \to \Xi_b \, \pi ]  ={\cal C}_{ab}\, \frac{4}{9}\, \frac{{\tilde g}^2}{2 \pi f^2} \,  \frac{m_{\Xi}}{m_{\Xi^{\prime *}}}|p_\pi|^3 \nn \\
\Gamma[\Xi_a^{\prime} \to \Xi_b^* \, \pi ]  &=&{\cal C}_{ab}\,\frac{8}{9}\, \frac{{\tilde g}^2}{2 \pi f^2} \,  \frac{m_{\Xi^*}}{m_{\Xi^{\prime}}}|p_\pi|^3 \qquad
\Gamma[\Xi_a^{\prime} \to \Xi_b \, \pi]  = {\cal C}_{ab}\,\frac{1}{9}\, \frac{{\tilde g}^2}{2 \pi f^2} \,  \frac{m_{\Xi}}{m_{\Xi^{\prime}}}|p_\pi|^3 \, .
\eea
Here ${\cal C}_{ab}$  is an $SU(3)$ factor which is $1/2$ for decays involving $\pi^0$ and one for decays involving
charged pions. The radially excited  doubly heavy strange baryons should also be below the
threshold for decays into kaons, and therefore should be quite narrow. The formulae in Eq.~(\ref{strong}) can be
used to obtain these decay widths as well. The isospin violating strong partial decay widths 
are obtained by using Eq.~(\ref{strong}) with  $C_{33}=\frac{2}{3}$ then multiplying by
$\theta^2$. The expected widths of these states are of order 10 keV, but more precise 
estimates cannot be made until the masses of the states and the coupling $\tilde g$ are known.
For the nonstrange doubly heavy  baryons, in the limit of infinite heavy quark mass, we obtain 
\bea 
\Gamma[\Xi^{\prime}] =\Gamma[\Xi^{\prime*}] = \frac{3 {\tilde g}^2}{4 \pi f^2} p_\pi^3  = 55 \, {\rm MeV} \, 
\left(\frac{{\tilde g}}{0.5}\right)^2 \,\left(\frac{p_\pi}{250 \, {\rm MeV}}\right)^3  
\eea 
for the total widths, and for the  branching fractions we find
\bea 
\frac{{\rm Br}[\Xi^{\prime *} \to \Xi^* \, \pi]}{{\rm Br}[\Xi^{\prime *} \to \Xi \, \pi]} = \frac{5}{4} \qquad 
\frac{{\rm Br}[\Xi^{\prime} \to \Xi^* \, \pi]}{{\rm Br}[\Xi^{\prime} \to \Xi \, \pi]} = 8 \, . 
\eea 
These relations  receive large corrections due to phase space effects.  Once the hyperfine splittings are taken into account the factors of $p_\pi^3$
will differ  greatly for the four decays. To get a feeling for these effects in the doubly charm sector we choose $m_{\Xi_{cc}} = 3440$ MeV,
$m_{\Xi^*_{cc}} = 3520$ MeV, $m_{\Xi^\prime_{cc}} = 3740$ MeV, and  $m_{\Xi^{*\prime}_{cc}} = 3820$ MeV, which corresponds to a diquark excitation
energy of 300 MeV and hyperfine splittings of 80 MeV. We then find
\bea 
\Gamma[\Xi_{cc}^{\prime}] =  {\tilde g}^2 \, 336 \, {\rm MeV} \qquad&& \Gamma[\Xi_{cc}^{*\prime}] = {\tilde g}^2 \, 78 \, {\rm
MeV} \nn \\ 
&&\nn \\  \frac{\Gamma[\Xi_{cc}^{\prime *} \to \Xi_{cc}^*  \, \pi]}{\Gamma[\Xi_{cc}^{\prime *} \to \Xi_{cc} \, \pi]} = 0.56  \qquad &&
\frac{\Gamma[\Xi_{cc}^{\prime} \to \Xi_{cc}^* \,  \pi]}{\Gamma[\Xi_{cc}^{\prime} \to \Xi_{cc} \,  \pi]} = 2.3 \, .
\eea 
Note that the $\Xi_{cc}^\prime$ unlike the $\Xi_{cc}^{\prime *}$ strongly prefers to decay to $\Xi_{cc}^*$ relative to $\Xi_{cc}$
despite the phase space suppression. This may be useful for distinguishing $\Xi_{cc}^{\prime *}$ and $\Xi_{cc}^{\prime}$
experimentally.

The SELEX $\Xi^{++}_{cc}(3780)$ is broad relative to the other SELEX doubly charm candidates.  Since it is 260 MeV heavier than the 
$\Xi^+_{cc}(3520)$, it is a natural candidate for one of the low lying excited doubly charm baryons. Unfortunately, no measurement of the 
width exists and the pattern of decays is also hard to understand, since  Ref.~\cite{Moinester:2002uw} states that 50\% of the decays 
to $\Lambda_c^+ K^- \pi^+\pi^+$ are through
$\Xi_{cc}^+(3520) \, \pi^+$ while the other 50\% are weak decays. More information on the quantum numbers of the $\Xi_{cc}^{++}(3780)$
and the $\Xi_{cc}^+(3520)$ are needed before we can determine which of the excited doubly charm baryons should be identified with the 
$\Xi_{cc}^{++}(3780)$.

In this paper we have developed a generalization of HH$\chi$PT which incorporates heavy quark-diquark symmetry and includes the leading symmetry
breaking corrections from the chromomagnetic couplings of the heavy quark and diquark.  We also included electromagnetic interactions in the Lagrangian,
and obtained an estimate of the width of the  $J=\frac{3}{2}$ member of the ground state doubly charm baryon doublet. The width of this state is
completely dominated by electromagnetic decays. We showed how to include the lowest lying doubly charm baryons which are expected to be excitations of
the doubly charm  diquark rather than the light degrees of freedom. Strong decay widths of low lying excited states were calculated and the states are expected
to be rather narrow because of limited phase space available for the decays. Of particular interest is the doubly charm strange sector where we expect
three pairs of excited baryons whose strong decay must violate isopsin conservation because they are below the kaon decay threshold. These states will
have narrow widths of 10 keV or less. Experimental efforts to observe the narrow doubly charm strange baryons would be of great interest.

\begin{acknowledgments}  

This research is supported in part by DOE grants DE-FG02-96ER40945 and DE-AC05-84ER40150
and an Outstanding Junior Investigator Award. 
 \end{acknowledgments}


\end{document}